\documentclass[12pt]{article}

\usepackage{dcolumn}

\usepackage{amssymb,amsmath,amsthm,bm,mathbbol}
\usepackage{graphicx,epstopdf,subfig}
\usepackage{xcolor,soul,color} 
\usepackage{times,authblk}

\begin{document}

\title{Maximum entropy and population heterogeneity in continuous cell cultures}

\author[1,2]{Jorge Fernandez-de-Cossio-Diaz\thanks{cossio@cim.sld.cu}}
\author[1,3]{Roberto Mulet\thanks{mulet@fisica.uh.cu}}
\affil[1]{Group of Complex Systems and Statistical Physics, Department of Theoretical Physics, University of Havana, Physics Faculty, Cuba}
\affil[2]{Systems Biology Department, Center of Molecular Immunology, Havana, Cuba}
\affil[3]{Group of Statistical Inference and Computational Biology, Italian Institute for Genomic Medicine, Italy }

\date{\today}

\maketitle

\begin{abstract}
Continuous cultures of mammalian cells are complex systems displaying hallmark phenomena of nonlinear dynamics, such as multi-stability, hysteresis, as well as sharp transitions between different metabolic states. In this context mathematical models may suggest control strategies to steer the system towards desired states. Although even clonal populations are known to exhibit cell-to-cell variability, most of the currently studied models assume that the population is homogeneous. To overcome this limitation, we use the maximum entropy principle to model the phenotypic distribution of cells in a chemostat as a function of the dilution rate. We consider the coupling between cell metabolism and extracellular variables describing the state of the bioreactor and take into account the impact of toxic byproduct accumulation on cell viability. We present a formal solution for the stationary state of the chemostat and show how to apply it in two examples. First, a simplified model of cell metabolism where the exact solution is tractable, and then a genome-scale metabolic network of the Chinese hamster ovary (CHO) cell line. Along the way we discuss several consequences of heterogeneity, such as: qualitative changes in the dynamical landscape of the system, increasing concentrations of byproducts that vanish in the homogeneous case, and larger population sizes.
\end{abstract}

\clearpage

\section*{Author Summary}
While the advantages of continuous culture in the biotechnological industry have been widely advocated in the literature, its adoption over batch or fed-batch modes stalls due to the complexities of these systems. In particular, continuous cell cultures display hallmark nonlinear phenomena such as multi-stability, hysteresis, and sharp transitions between metabolic phenotypes. Moreover, the impact of the heterogeneity of a cell population on these features is not well understood. We use the maximum entropy principle to model the phenotypic distribution of an heterogeneous population of cells in a chemostat. Given the metabolic network and the dilution rate, we obtain a self-consistent solution for the stationary distribution of metabolic fluxes in cells. We apply the formalism in two examples: a simplified model where the exact solution is tractable, and a genome-scale metabolic network of the Chinese hamster ovary (CHO) cell line widely used in industry. We demonstrate that heterogeneity may be responsible for qualitative changes in the dynamical landscape of the system, like the disappeareance of a bistable regime, the increase of concentrations of byproducts that vanish in the homogeneous system and larger number of cells. We explain the causes behind these phenomena.
\clearpage

\section*{Introduction}

Recombinant protein production requires suitable cell hosts and culture conditions \cite{wurm2004production}. For this purpose mammalian cells are often grown in chemostat-like cultures where a continuous flow of incoming fresh media replaces culture liquid containing cells and metabolites. Alternative processes such as batch or fed-batch are also adopted by many industrial facilities, but the advantages of the continuous mode have been predicted to drive its wide adoption in the near future \cite{werner1992safety, griffiths1992animal, kadouri1997some, werner1998letter, croughan2015future, konstantinov2015white}. However, experiments have demonstrated that continuous cultures exhibit hallmark phenomena of nonlinear dynamics, such as multiple steady states under identical external conditions \cite{europa2000multiple, altamirano2001analysis, hayter1992glucose, gambhir2003analysis} and hysteresis loops \cite{follstad1999metabolic, europa2000multiple, cossio2017ploscb}. Sophisticated control strategies are then required to drive the system towards desired steady states.

In this context, mathematical modeling has been used with some success \cite{cossio2017ploscb, yongky2015mechanism, smith1995theory}, but most of these works deal with simple cell populations, consisting at most of a few competing species \cite{smith1995theory,Lee1991Application}. Although it is known that no two cells in culture are alike \cite{kiviet2014stochasticity}, the effects of individual cell-to-cell variability are seldom considered \cite{delvigne2014variability}. Attempts to model heterogeneity in cell cultures are often based in population balance models \cite{Henson2003Dynamic} or similar approaches, which require prior postulation of the mechanism driving the heterogeneity and depend on more quantitative parameters than homogeneous modeling. These models are affected in part by the limited availability of quantitative data \cite{gonzalez2017heterogeneity}, but also by an incomplete understanding of the role played by different mechanisms driving heterogeneity. Indeed, many complex processes contribute to heterogeneity in a cell population, including gene expression noise \cite{elowitz2002stochastic}, partition errors at cell division \cite{fernandez2018cell, huh2011random}, mutations \cite{wang2016clonal}, size variability \cite{tzur2009cell}, as well as environmental gradients in the culture \cite{lara2006living}.

It is then important to understand what features change or are missed in a model of continuous culture by treating the cell population as an homogeneous system. In this work we propose to apply the maximum entropy principle \cite{jaynes1957information,jaynes2003ptlos, harte2014maxentecology, schneidman2006nature, martino2017statistical,de2016growth,de2017quantifying} to model cell heterogeneity in a continuous bioreactor. Although we build our theory based on the model of a chemostat presented in \cite{cossio2017ploscb}, and discuss specific metabolic networks, the framework can accommodate easily other models of continuous cell cultures and metabolic networks.

\section*{Materials and Methods}

\subsection*{General framework}

We study the steady states of a population of cells inside a well-mixed bioreactor, where fresh medium continuously replaces culture fluid at a given dilution rate $D$. Each cell will be described by vectors $\underline r^{(h)}$ and $\underline u^{(h)}$, giving the flux of every reaction in the metabolic network of the cell line under study and the metabolite uptake rates, respectively. Here and in what follows, the super-index $h$ denotes an individual cell. If $N_{ik}$ denotes the stoichiometric coefficient of metabolite $i$ in reaction $k$ ($N_{ik} > 0$ if metabolite $i$ is produced in the reaction, $N_{ik}<0$ if it is consumed, and $N_{ik} = 0$ if it does not participate in the reaction), then cell $h$ produces metabolite $i$ at an overall rate $\sum_k N_{ik} r_k^{(h)}$. This production must balance the cellular demands for metabolite $i$. In particular we consider a constant maintenance demand $e_i^{(h)}$ which is independent of growth \cite{kilburn1969energetics, sheikh2005modeling}, and the requirements $y_i^{(h)}$ of each metabolite for the synthesis of biomass \cite{feist2010biomass, feist2016cells}. If biomass is synthesized at a rate $z^{(h)}$, we obtain the following overall balance equation for each metabolite $i$:
\begin{equation}
\sum_k N_{ik} r_k^{(h)} + u_i^{(h)} = e_i^{(h)} + y_i^{(h)} z^{(h)}
\label{eq:balance}
\end{equation}
For simplicity we will assume that the overall macromolecular composition of the cell is constant, \emph{i.e.} that $e_i^{(h)} = e_i$ and $y_i^{(h)}=y_i$ are independent of $h$. The net growth rate of a cell depends both on the rate of biomass synthesis, $z$, and on the concentration of metabolites in the media. We assume that some metabolites are toxic (such as lactate and ammonia) and affect the growth rate of cells. In order to accommodate the most common dependencies found in the literature, we assume:
\begin{equation}
\lambda^{(h)} = z^{(h)} \times K(\underline s) - \sigma(\underline s)
\label{eq:growth-rate}
\end{equation}
where $\underline s$ is the vector of metabolite concentrations in the culture, $\lambda^{(h)}$ is the net growth rate, $K(\underline s)$ is a growth inhibition factor, and $\sigma(\underline s)$ a death rate.

In addition to Equation \ref{eq:balance}, the fluxes $r_k^{(h)},u_i^{(h)}$ must respect physico-chemical constrains arising from thermodynamics and molecular crowding. These constrains will also depend on the extracellular conditions of the culture, specifically the concentrations $\underline s$ of metabolites. They can be summarized as:
\begin{equation}
\sum_k \alpha_k |r_k^{(h)}| \le C
\label{eq:flux-costs}
\end{equation}
\begin{equation}
\mathrm{lb}_k \le r_k^{(h)} \le \mathrm{ub}_k
\label{eq:flux-bounds}
\end{equation}
\begin{equation}
-L_i \le u_i^{(h)} \le U_i
\label{eq:ui-inequality}
\end{equation}
where $\alpha_k$ are the enzymatic flux costs per unit of flux through reaction $k$, $C$ is the maximum enzymatic cost available to the cell, $\mathrm{lb}_k,\mathrm{ub}_k$ are the lower and upper bounds of reaction $k$, $L_i = 0$ for metabolites that cannot be excreted from the cell and $L_i = \infty$ otherwise, $U_i$ is the maximum uptake rate of metabolite $i$, and $X$ the total number of alive cells. The value of $U_i$ will depend on the conditions of the culture and will be specified below. The reaction bounds $\mathrm{lb}_k,\mathrm{ub}_k$ are $-\infty$ and $\infty$ respectively for reversible reactions, while $\mathrm{lb}_k=0,\mathrm{ub}_k=\infty$ for irreversible reactions.

The constrains (\ref{eq:balance}--\ref{eq:ui-inequality}) define a convex polytope of feasible metabolic states \cite{palsson2015book,Gerstl2018Fluxtope} that we denote by $\mathcal P_{\underline s}$. Each point within this space consists of coordinates $\underline v = (\underline r, \underline u)$ which fully specify the metabolic state of a cell in the model. Let $P_{\underline s}(\underline v)$ be the density of cells with metabolic fluxes $\underline v$. To determine the form of $P_{\underline s}(\underline v)\mathrm d\underline v$, we adopt the Principle of Maximum Entropy (MaxEnt), which in this context can be stated as follows \cite{martino2017statistical,jaynes1957information}:
\begin{quote}
    Given the set of allowed metabolic states ($\mathcal P_{\underline s}$), the dependency of the cellular growth rate with metabolic fluxes ($\lambda_{\underline s}(\underline v)$), and the average growth rate of the cell population $\langle\lambda\rangle$, the distribution of cells within $\mathcal P_{\underline s}$ has the form $P_{\underline s}(\underline v) \propto \mathrm e^{\beta \lambda_{\underline s}(\underline v)}$, where $\beta$ is chosen so that the expectation of $\lambda$ under $P_{\underline s}(\underline v)$ coincides with $\langle\lambda\rangle$.
\end{quote}
For $\beta=0$ every point in the space $\mathcal P_{\underline s}$ is equally likely. In this case cells explore uniformly the space of allowed solutions defined by Equations (\ref{eq:balance}-\ref{eq:ui-inequality}). For $\beta=\infty$ the distribution is a Dirac delta with infinite mass on the points of maximum growth rate. In the language of Statistical Mechanics this is the ground state of the system, also known in the Systems Biology literature as the Flux Balance Analysis (FBA) solution \cite{palsson2015book}.

\subsection*{Simple metabolic model}

\begin{figure}
    \centering
    \includegraphics[width=5cm]{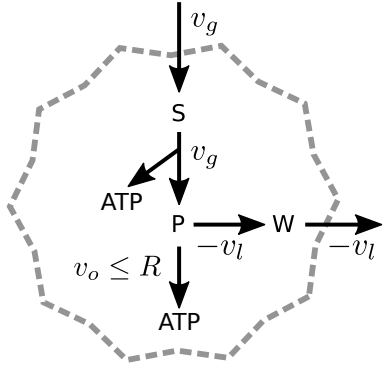}
    \caption{\label{fig:0} \textbf{Simplest metabolic network exhibiting a switch} A primary carbon source (glucose) is consumed at a rate $v_g$. It is processed into an intermediate P, generating energy (ATP). The intermediate can be excreted generating waste (W) at a rate $-v_l$, or it can be further oxidized via the respiratory pathways (rate $v_o$). The respiratory pathway is capped, $v_o \le R$.}
\end{figure}

To gain some insight, we analyze first a simplified metabolic model that exhibits a switch from oxidative pathways at low growth rates to fermentative pathways at higher growth rates \cite{vazquez2010catabolic}. The model consists of the following stoichiometric constrains:
\begin{equation}
2v_g + v_l - v_o = 0
\label{eq:toy-sto}
\end{equation}
\begin{equation}
2v_g + 18 v_o = v_{atp} 
\label{eq:toy_vatp}
\end{equation}
\begin{equation}
v_g \le \min\{V_g, c_g D/X\}
\label{eq:toy_vg}
\end{equation}
\begin{equation}
v_l \le 0, 0 \le v_o \le R
\label{eq:toy_vl}
\end{equation}
where $v_g$ is the velocity of consumption of glucose, $v_l$ the excretion of lactate, $v_o$ the velocity of oxidative phosphorylation, and $v_{atp}$ the synthesis of ATP (see Figure \ref{fig:0}). We only consider the ATP requirements of growth and maintenance and ignore additional biomass components. Therefore the rate of biomass synthesis ($z$) is an affine function of the rate of ATP synthesis, $z = (v_{atp} - e)/y$, where $y$ is the biomass yield and $e$ is the constant maintenance demand for ATP \cite{kilburn1969energetics}. The rate of oxidative phosphorylation is caped by the limited capacity of mitochondria, $R$ \cite{cossio2017limits}. This simplified model or close variations have been extensively discussed in the literature, \emph{e.g.} \cite{cossio2017microenvironmental, vazquez2010catabolic, cossio2017ploscb, molenaar2009shifts}.

We assume that the secreted lactate is toxic, so that the net growth rate is given by $\lambda = z - \sigma$, where the death rate $\sigma$ is some increasing function of the extracellular concentration of lactate, $w$. For simplicity we use a linear form, $\sigma=\tau w$, where $\tau$ is the death rate per unit of lactate. One can think of $\tau$ as the slope of a more realistic description of the dependence of the death rate on the concentrations of toxic compounds, thus avoiding the introduction of too many parameters into this simplified model.

Parameters are based on experimental values obtained for mammalian cells in the literature. The ATP maintenance demand $e = 1\mathrm{mmol/gDW/h}$ has been measured for mouse LS cells \cite{kilburn1969energetics} (see \cite{lynch2015bioenergetic} for similar values in other cell types). The value $R = 0.45 \text{mmol/gDW/h}$ was calculated in \cite{cossio2017ploscb}, based on measurements of the glucose uptake threshold that triggers a switch to fermentation in mammalian cells \cite{vazquez2010catabolic}. The maximum velocity of glucose uptake, $V_g = 0.5 \text{mmol/gDW/h}$ is based on a value measured for HeLa cells \cite{rodriguez2009kinetics}. The value $y=348 \text{mmol/gDW}$ was then adjusted so that the maximum growth rate matches typical duplication rates of mammalian cells of $~1/\mathrm{day}$. From a linear approximation of the death rate dependence on lactate concentration measured in a mammalian culture \cite{dhir2000bb} we obtain $\tau = 0.0022\mathrm{h}^{-1}\mathrm{mM}^{-1}$. The concentration of glucose was set at a value typical of mammalian cell culture media, $c_g = 15 \text{mM}$.

\subsection*{Metabolic network of CHO-K1}

Motivated by the fact that most therapeutic proteins requiring complex post-translational modifications are produced in Chinese hamster ovary (CHO) host cell-lines \cite{wurm2004production}, our second example is a metabolic model of this cell line. We employed a CHO-K1 line-specific metabolic model, based on the latest reconstruction of CHO metabolism available at the time of writing \cite{hefzi2016cho}. The network recapitulates experimental growth rates, essential enzyme requirements and cell line specific amino acid auxotrophies. In order to enforce \eqref{eq:flux-costs}, we complemented this network with a set of reaction costs of the form $\alpha_i = MW_i / a_i$, where $MW_i$ is the molecular weight of the enzyme catalyzing reaction $i$ and $a_i$ the specific activity. These values were mined by T. Shlomi \emph{et. al} \cite{shlomi2011ploscb} from public enzyme data repositories. For reactions where the corresponding value could not be found, we set its flux cost to the median of all the flux costs available. Finally $C = 0.078\mathrm{mg/mgDW}$ is the mass fraction of metabolic enzymes in the dry weight of a typical mammalian cells, which can be estimated from protein abundance measurements \cite{Vogel2010Sequence,shlomi2011ploscb}. The predictions of combining flux balance analysis with a crowding constrain obtained in this manner have been shown to exhibit significant correlation with enzyme mRNA expression levels in mammalian cells \cite{shlomi2011ploscb}.

A constant maintenance demand was added as an ATP hydrolysis flux at a rate $1\text{mmol/gDW/h}$, based on measurements for mouse LS cells \cite{kilburn1969energetics}. The maximum glucose uptake was set at $V_g = 0.5 \text{mmol/gDW/h}$, a value measured in HeLa cells \cite{rodriguez2009kinetics}. Unfortunately kinetic parmaeters to estimate $V_i$ for many other metabolites are not available. However, based on multiple reports in the literature \cite{ozturk1992nh4,dhir2000bb,altamirano2001improvement} we estimated that amino acid uptakes are typically tenfold slower than sugar uptake, and therefore set $V_i = V_g/10$ for amino acids. In the simulations we used Iscove's modified Dulbecco's medium (IMDM) to set the values of $c_i$, and set $V_i = \infty$ and infinite concentrations for water, protons and oxygen (see Supplementary Materials for full specification). Since lactate and ammonia are the most commonly recognized toxic byproducts in mammalian cell cultures, we set $\lambda = K \times z$ with
\begin{equation}
K = (1 + s_\mathrm{nh4}/K_\mathrm{nh4})^{-1} (1 + s_\mathrm{lac}/K_\mathrm{lac})^{-1} 
\end{equation}
with $K_\mathrm{nh4} = 1.05\text{mM}$ and $K_\mathrm{lac} = 8\text{mM}$, based on the values and functional form reported by \cite{bree1988bb}.

\subsection*{Expectation propagation approximation for the computation of moments}

In order to compute the steady state concentrations $s_i$, it is necessary to compute the expected value of $u_i$ under the MaxEnt distribution. Although in the simple model this poses no difficulty, in a genome-scale metabolic network such as the CHO-K1 considered in this paper the vector $\underline v$ has hundreds of components and an exact computation \cite{avis1990lrs} becomes intractable. A \emph{hit and run} Monte Carlo approach has been used for moderately sized networks \cite{martino2017statistical,Martino2015Uniform}. Although these methods are guaranteed to converge to an uniform sample, this is only true in the asymptotic limit of an infinite number of steps. Unfortunately the geometrical shape of the metabolic flux space is highly elongated in some directions but very compressed in others \cite{Martino2015Uniform}. In practice it becomes very hard to determine how long the Monte Carlo computation should run to achieve convergence, particularly so for large metabolic networks.

A better approach is to use message-passing algorithms \cite{fernandez2016fast}. Recently, Expectation Propagation (EP) \cite{minka2001ep} has been successfully employed to compute a very good approximation of the marginal flux distributions in absence of selection ($\beta = 0$) \cite{braunstein2017natcomm}. In \cite{braunstein2017natcomm} the reader can find an exhaustive assessment of the quality of this approximation in a variety of metabolic networks. In Supplementary Materials we describe how the same method can be used to approximate the marginal flux distributions for non-zero $\beta$.

\subsection*{Additional details}

All model simulations were carried out in Julia \cite{bezanson2017julia}. The CHO-K1 metabolic network \cite{hefzi2016cho} was loaded and setup using the {COBRApy} package \cite{ebrahim2013cobrapy}.
The expectation propagation implementation was taken from https://github.com/anna-pa-m/Metabolic-EP \cite{braunstein2017natcomm}.

Since taking the absolute value in \eqref{eq:flux-costs} is not a linear operation, we must replace reversible reactions in the model by two reactions, one in the forward and another in the backward direction. This ensures that all reaction fluxes are non-negative variables and \eqref{eq:flux-costs} becomes a linear inequality. However, this almost doubles the number of reactions in the CHO-K1 model, which makes Expectation Propagation extremely slow to converge or in some cases it even fails to do so. In order to obtain a reduced tractable network, we first solved our model at $\beta=\infty$, which can be done with standard linear programming packages \cite{cossio2017ploscb}. Then we removed from the CHO-K1 model all reactions that were identically zero for all values of $\xi$. This results in a reduced model with 380 metabolites and 401 reactions (provided as supplementary materials), where Expectation Propagation converges robustly and fast.

\section*{Results}

\subsection*{General solution of the model}

In order to determine $P_{\underline s}$ we need to know the concentrations $s_i$, the function $\lambda_{\underline s}(\underline v)$ giving the growth rate for each feasible metabolic state, and the average growth rate $\langle \lambda\rangle$. The later is easy in the chemostat: $\langle\lambda\rangle = D$ (the dilution rate) \cite{smith1995theory}. It is much more difficult to obtain experimental data for all the relevant $s_i$. In the worst case in which no information about the $s_i$ is available, we notice that when the chemostat is in steady state, the input flux of metabolite $i$ must balance its consumption by the cells and its output flow. The input and output fluxes per culture volume in the chemostat are given by $D c_i$ and $D s_i$, respectively, where $c_i$ is the concentration of metabolite $i$ in the external feed. Since the rate of consumption of individual cells is $u_i^{(h)}$, the total consumption by the population of cells can be estimated as:
\begin{equation}
    \sum_h u_i^{(h)} \approx X \int u_i(\underline v) P_{\underline s}(\underline v) \mathrm d \underline v
\end{equation}
where $X$ is the total number of cells. In the limit of a large number of cells this equation becomes exact. In steady state, the concentration $s_i$ is constant, which leads to the following mass-balance equation:
\begin{equation}
    D(c_i - s_i) - X \int u_i(\underline v) P_{\underline s}(\underline v) \mathrm d \underline v = 0
\label{eq:si-balance}
\end{equation}
To obtain the steady state values of $s_i$, equation \eqref{eq:si-balance} must be solved self-consistently together with the MaxEnt form of the distribution $P_{\underline s}(\underline v)$. Moreover since the steady state value of $s_i$ must be non-negative, we can extract from this equation an approximate form for the maximum uptake rate that further simplifies the computation (see Supplementary Materials for a derivation of this equation from Michaelis-Menten kinetics),
\begin{equation}
U_i = \min\{V_i,c_i X/D\} = \min\{V_i,c_i\xi\}
\label{eq:Ui}
\end{equation}
where $V_i$ is the maximum uptake rate of metabolite $i$. The solution of the model in the case $\beta = \infty$ (FBA) is described in detail in Ref. \cite{cossio2017ploscb}, where it was argued that the ratio $\xi = X/D$ determines the steady state of the culture and links steady states in the chemostat with those in perfusion systems. It has units (cells $\times$ time / volume), and can be interpreted as the number of cells maintained alive for a period of time per unit of volume of fresh media supplied into the culture.

In the case of finite $\beta$, the value of $\xi$ only determines the shape of the polytope of phenotypic states that cells can adopt. According to the maximum entropy principle, the distribution of cells within this polytope is of the form:
\begin{equation}
    P(\underline v) = \frac{\mathrm e^{\beta^\prime z(\underline v)}}{\int_{\mathcal P}\mathrm e^{\beta^\prime z(\underline v)}\mathrm d\underline v}
    \label{eq:maxentP}
\end{equation}
where $\beta^\prime = \beta K(\underline s)$ and the parameter $\beta$ quantifies the level of heterogeneity in the population of cells. It is a macroscopic representation of the underlying noisy processes that sustains cell-to-cell variability in the population. Small values of $\beta$ lead to an almost uniform distribution $P(\underline v)$ over all possible states. This corresponds to a highly heterogeneous population. In contrast, at larger values of $\beta$ the population concentrates around the FBA solution maximizing the growth rate. This corresponds to a highly homogenous population.

From \eqref{eq:maxentP}, we compute the expected values of the exchange fluxes $\langle u_i\rangle$ using the Expectation Propagation algorithm described in the appendix \cite{braunstein2017natcomm}. Next, the values of the metabolite concentrations are obtained from \eqref{eq:si-balance}:
\begin{equation}
    s_i = c_i - \langle u_i\rangle \xi
    \label{eq:si-solution}
\end{equation}
Given $\underline s$, we compute $K(\underline s),\sigma(\underline s)$ and then $\beta = \beta^\prime / K(\underline s)$. Similarly, $\langle \lambda\rangle = \langle z\rangle K(\underline s) - \sigma(\underline s)$, which then determines the value of the dilution rate consistent with this solution, $D = \langle \lambda\rangle$. Finally, the total number of cells in the steady state is given by $X = \xi D$.


\subsection*{Simple metabolic model}

Within our framework the simple metabolic model admits an exact solution that provides important clues about the role of heterogeneity in more realistic models.
For a given value of $\xi$, in this case the MaxEnt distribution takes the form:
\begin{equation}
    P(v_{atp}, v_g; \xi) = \mathrm{e}^{-\beta v_{atp}/y} / Z(\xi)
    \label{eq:Ptoy}
\end{equation}
for $(v_{atp},v_g)\in \mathcal P_\xi$, where $\mathcal P_\xi$ is the polytope defined by the constrains \eqref{eq:toy-sto}--\eqref{eq:toy_vl} (after eliminating $v_l,v_o$), and
\begin{equation}
    Z(\xi) = \int_{\mathcal P_\xi} \mathrm{d}v_{atp}\mathrm{d}v_g \mathrm{e}^{-\beta v_{atp}/y}
\end{equation}
Notice that constant terms, including the additive death rate $\sigma$ and the maintenance demands, cancel upon normalization. Due to the low-dimensionality of this model, the moments of \eqref{eq:Ptoy} can be evaluated by numerical integration to any desired accuracy. Then the steady state concentrations of glucose and lactate can be calculated using \eqref{eq:si-solution}.

\begin{figure}
    \centering
    \includegraphics[width=15cm]{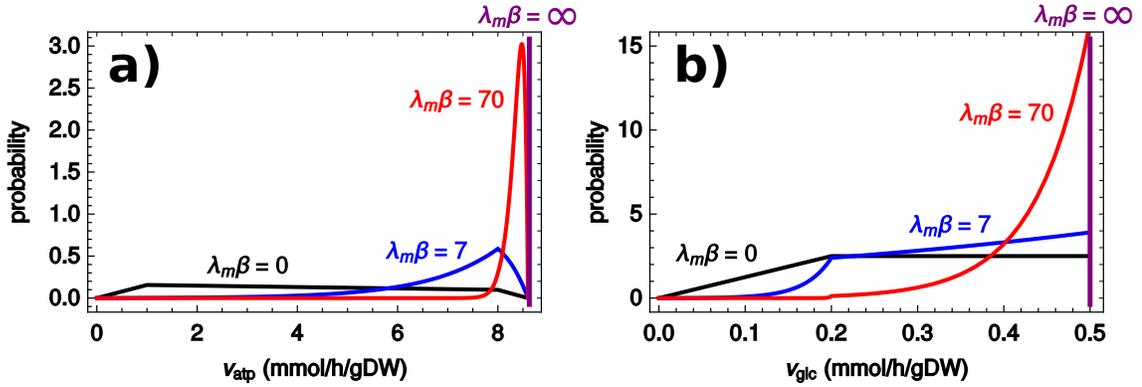}
    \caption{\label{fig:toy-vatp} \textbf{Distributions of the fluxes $v_{atp}$ and $v_{g}$.} Distribution among cells of the glycolytic flux ($v_g$) and the ATP synthesis reaction flux ($v_{atp}$). Since $\beta$ has units of inverse time, we use the dimensionless parameter $\lambda_m\beta$, where $\lambda_m$ is the maximum growth rate at $\xi = 0, \beta=\infty$.}
\end{figure}

Figure \ref{fig:toy-vatp} shows typical flux distributions of $v_{atp}$ and $v_{glc}$ in the population of cells, for different values of $\xi$ and $\beta$. As $\beta$ increases (more homogeneous populations), cells cluster around the maximum feasible rate of ATP production, which also coincides with maximum glucose consumption. In the $\beta\rightarrow\infty$ limit (FBA) the distribution becomes a Dirac delta (purple line) localized in this optimal point. On the other hand, when $\beta\rightarrow0$ cells distribute uniformly in the space of feasible metabolic states, favoring states with low growth rate. This observation has been interpreted as implying that higher growth rates require active regulation in the cell \cite{martino2017statistical} (see Supplementary Material for a study of an evolutionary model consistent with this explanation within our framework).

\begin{figure}
    \centering
    \includegraphics[width=15cm]{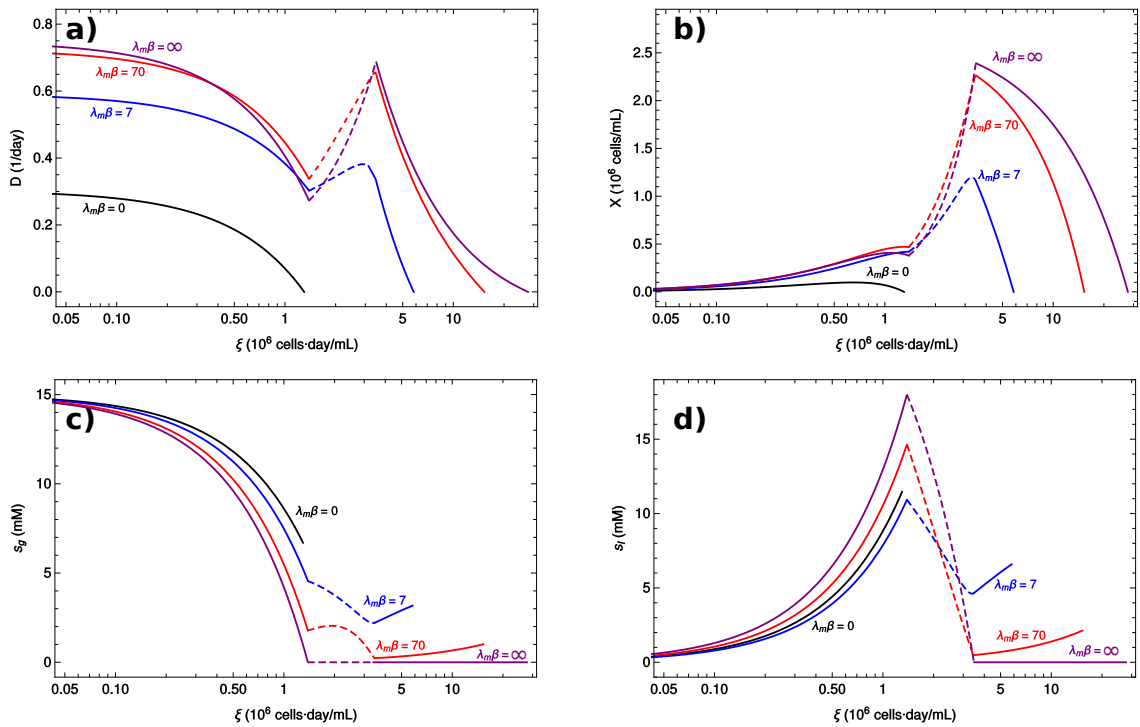}
    \caption{\label{fig:toy-xi} \textbf{Steady states of the simple metabolic network as a function of $\xi$.} The different panels show the steady state values of $D$, $X$, $s_g$ and $s_w$, as functions of $\xi$ for different values of $\beta$ (indicated in the legend). Discontinuous lines indicate unstable steady states.}
\end{figure}

Figure \ref{fig:toy-xi} shows the solution of the model as a function of $\xi$, for selected representative values of $\beta$. For comparison the $\beta = \infty$ solution is shown in purple. As the crossing curves in Fig. \ref{fig:toy-xi}a indicate, the effect of decreasing $\beta$ is not simply to decrease the average growth rate of the population of cells. At first sight this seems to contradict the fact that the $\beta = \infty$ solution is where cells adopt a phenotype with maximum growth rate (the FBA solution). However, due to the accumulation of toxic byproducts in the culture, maximizing $z$ may result in higher toxicities, and the net effect is to decrease the overall growth rate of the population. This translates into the fact that the heterogenous population may have higher cell numbers than the homogenous one (where the red curve is higher than the purple in Fig. \ref{fig:toy-xi}b). This explanation is confirmed in Fig. \ref{fig:toy-xi}d, which shows that the concentration of lactate reaches the highest levels when $\beta = \infty$.

A second feature connected to heterogeneity is that even for high values of $\beta$ and $\xi$, the concentration of lactate in the culture increases with $\xi$. However in the strictly homogeneous limit ($\beta=\infty$) it goes to zero. This striking difference has important implications in the interpretations of bulk metabolic measurements in populations of cells. For example, the observation of an increasing concentration of lactate could be interpreted as the result of a selective pressure, pushing cells towards fermentation. On the contrary, our model shows that in a chemostat this is a natural consequence of the heterogeneity of the population.

\begin{figure}
    \centering
    \includegraphics[width=8cm]{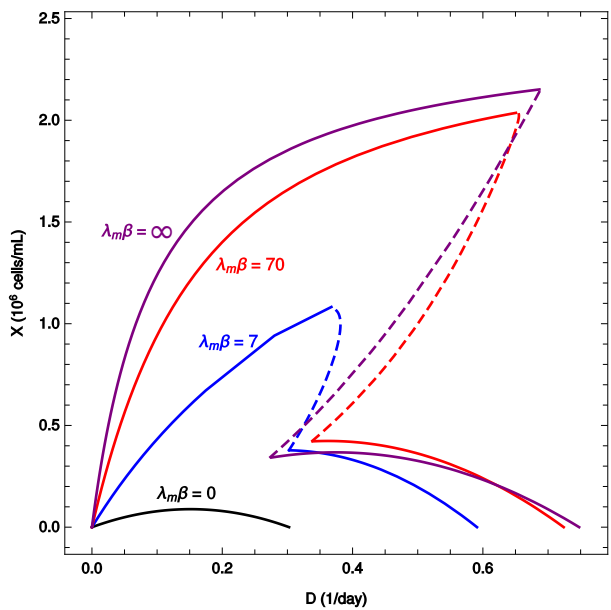}
    \caption{\label{fig:3} \textbf{Cell density versus dilution rate in the simple metabolic model.} Plot of the steady state cell density as a multi-function of the dilution rate, for different values of $\beta$. Discontinuous lines indicate unstable steady states.}
\end{figure}

Moreover notice that, for high enough $\beta$, the curve of $D$ versus $\xi$ (Fig. \ref{fig:toy-xi}a) displays multistability. To the same value of $D$ there may correspond more than one value of $\xi$. A theorem proved in Ref. \cite{cossio2017ploscb} establishes that a steady state is stable if $D(\xi)$ is decreasing at $\xi$. The theorem also holds in the present model, where the heterogeneity only redefines the function $D(\xi)$ to which the theorem was originally applied. In this case, an important consequence of the heterogeneity is that it reshapes the dynamical landscape of the system, which is tightly connected to the decrease of the accumulation of lactate in the system as $\beta$ decreases (cf. Fig. \ref{fig:toy-xi}d). For example, increasing the heterogeneity (decreasing $\beta$) abolishes the bistable regime. This is shown clearly in Figure \ref{fig:3}, where the steady state values of $X$ are plotted against the dilution rate.

\begin{figure}
    \centering
    \includegraphics[width=15cm]{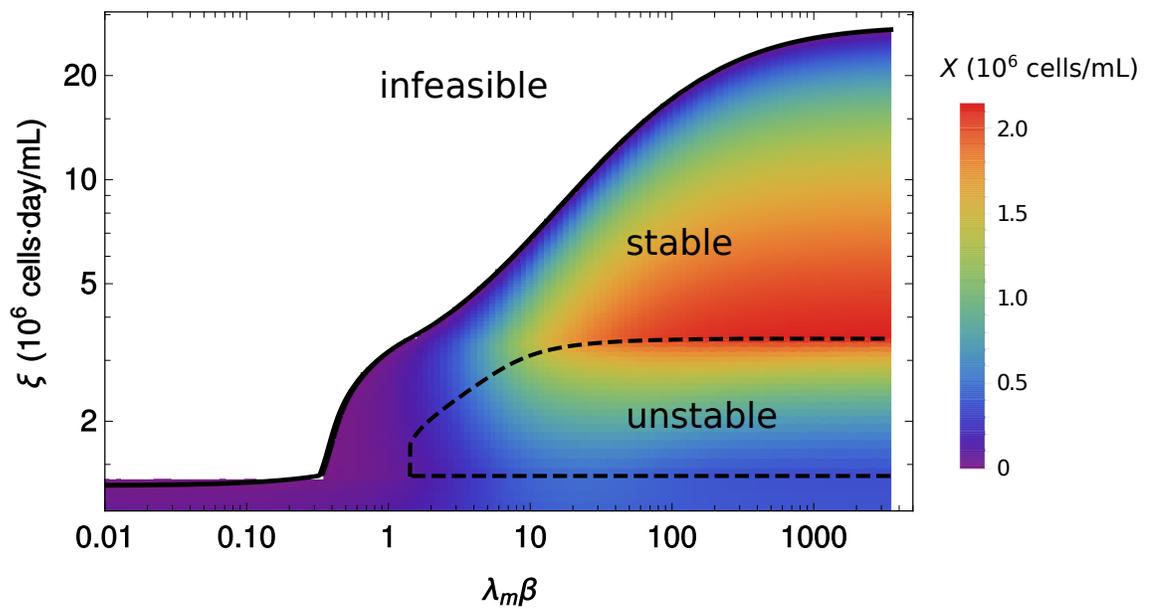}
    \caption{\label{fig:beta-xi} \textbf{Critical values of $\xi$ versus $\beta$.} Medium depth (continuous line) and critical values of $\xi$ separating stable steady states from unstable steady states (discontinuous lines), versus $\beta$. The gradient shows the steady state concentration of cells ($X$).}
\end{figure}

Heterogeneity also has the undesirable consequence of decreasing the \emph{medium depth} ($\xi_m$), the maximum value of $\xi$ with a non-zero steady state concentration of live cells. It is important to realize that even if the heterogeneity may help to increase the number of cells in the presence of toxicity (as discussed in the previous paragraphs), the medium depth never decreases with $\beta$. A plot of $\xi_m$ as a function of $\beta$ is shown in Figure \ref{fig:beta-xi} (continuous line). For highly heterogeneous systems (low values of $\beta$), $\xi_m$ is almost insensitive to changes in $\beta$. Then there is a sharp slope change after which $\xi_m$ steadily increases with $\beta$.

Since the stability of the system depends on the level of heterogeneity, in Figure \ref{fig:beta-xi} we show how the values of $\xi$ corresponding to unstable steady states depend on $\beta$ (discontinuous lines). This way the $\xi,\beta$ plane is divided into three regions: \emph{infeasible}, where no steady state is possible, \emph{stable} and \emph{unstable}, according to the type of steady state. This confirms the dramatic effect of heterogeneity on the dynamic landscape of the system, as unstable steady states disappear below a certain threshold value of $\beta$, \emph{i.e.} after a critical level of heterogeneity. For large $\beta$, the system becomes more robust in the sense that the range of values of $\xi$ defining unstable states is constant, while the stable regime becomes wider. 

Figure \ref{fig:beta-xi} also shows the density of cells in steady state for each pair of values $(\beta,\xi)$ (color gradient). Notice that in this model, the highest cell densities occur near unstable states. This should be interpreted as a word of caution, since trying to increase the number of cells in the industrial setting can have the undesired effect of washing the culture.

\subsection*{Analysis of a genome-scale metabolic network of the CHO-K1 cell line}

\begin{figure}
    \centering
    \includegraphics[width=14cm]{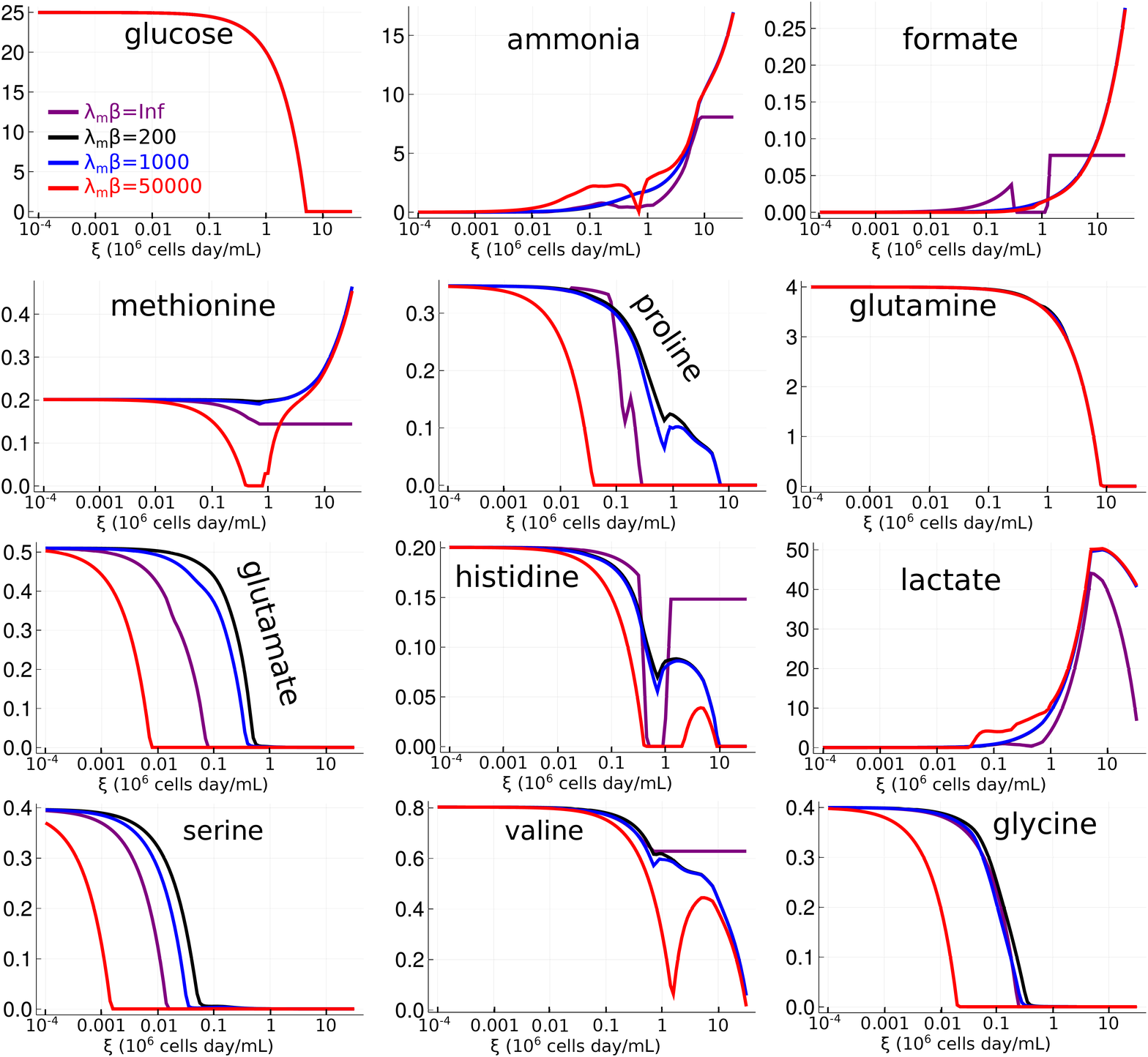}
    \caption{\label{fig:cho-conc} \textbf{Steady state metabolite concentrations as functions of $\xi$ for the CHO-K1 model.} Steady state concentrations of selected metabolites as functions of $\xi$, for the simulations of the CHO-K1 metabolic network. Representative values of $\beta$ are plotted (see legend). The purple line corresponds to the $\beta\rightarrow \infty$ limit. To reduce clutter, this figure does not distinguish between stable and unstable steady states.}
\end{figure}

Next, we study a reconstruction of the CHO-K1 cell line metabolic network \cite{hefzi2016cho}. Figure \ref{fig:cho-conc} shows the steady state concentrations of selected metabolites as functions of $\xi$ and for certain representative values of $\beta$. There are several differences between the homogeneous (shown in purple in the plots) and heterogeneous regimes. Formate, a byproduct of mitochondrial oxidative metabolism secreted by normal tissues and especially cancer cells \cite{meiser2018formateoverflow}, stabilizes at a constant concentration for large $\xi$ when $\beta=\infty$. The selective pressure for growth predicted by an FBA calculation only favors a mild secretion of this metabolite that is not enough to support its accumulation at larger $\xi$. In contrast, even mild levels of heterogeneity result in an increasing accumulation of formate at larger values of $\xi$. Ammonia shows a similar behavior. An heterogeneous population shows an accumulation of formate and ammonia that does not originate from the selective pressures for faster growth acting on individual cells. These differences resemble the behavior of lactate in the simple model.

In the case of histidine and valine, we have the opposite situation, where the homogeneous model predicts non-vanishing concentrations at large $\xi$, but even mild levels of heterogeneity drive the concentrations of these metabolites to zero. Other metabolites show less significant differences, such as glutamate, serine and glycine, where the level of heterogeneity only controls the rate of depletion as $\xi$ increases but does not seem to produce any qualitative differences. Glucose and glutamine are almost insensitive to variations in $\beta$.
This means that the structure of the metabolic network itself favors maximal consumption of these metabolites, even in absence of active regulation.

\begin{figure}
    \centering
    \includegraphics[width=14cm]{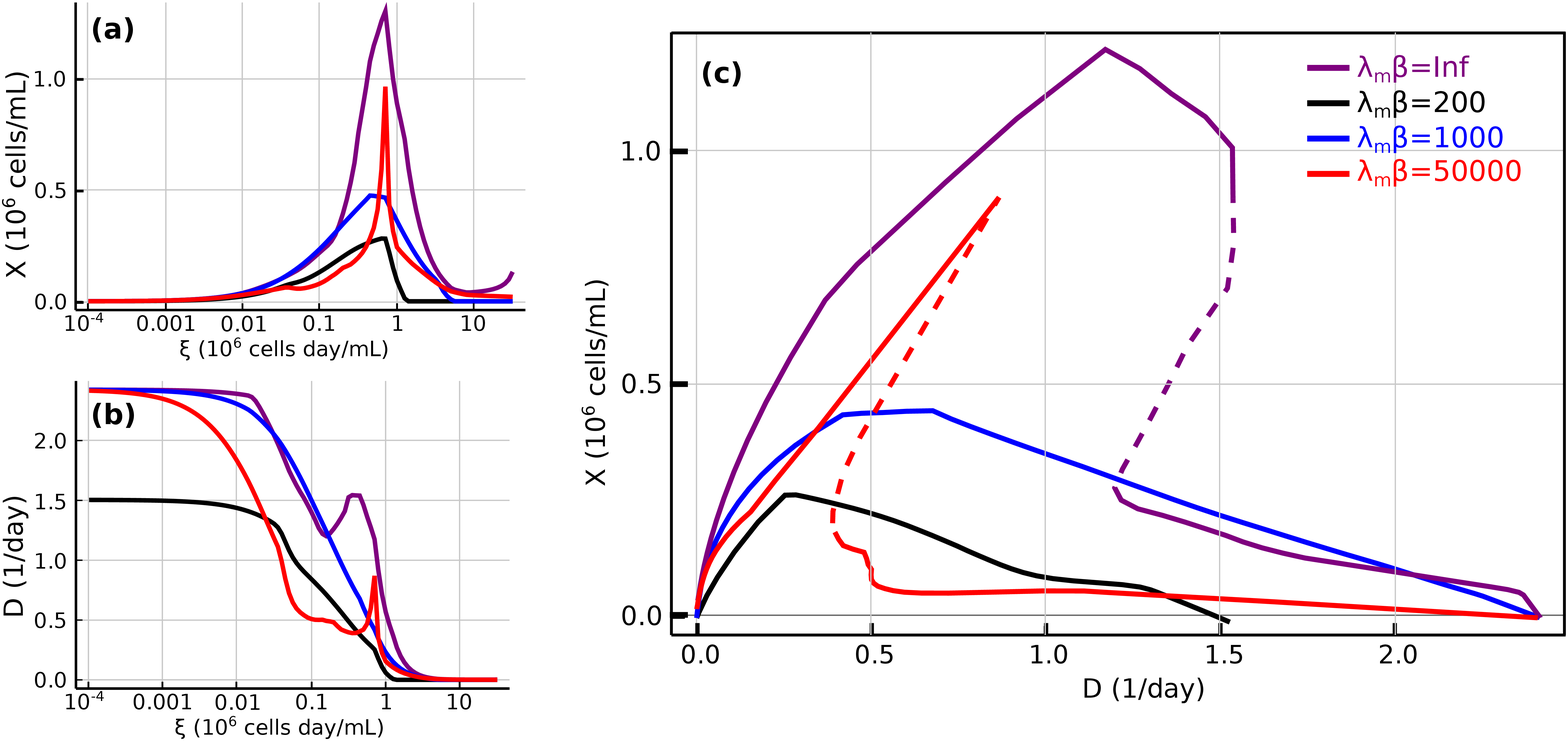}
    \caption{\label{fig:X-D-cho} \textbf{Dilution rate and cell density in steady states of the CHO-K1 metabolic model.} Different curves correspond to different levels of heterogeneity. In (c), discontinuous line indicates unstable steady states.}
\end{figure}

In Figure \ref{fig:X-D-cho}(a,b) we plot the dilution rate and the cell concentration in steady state as functions of $\xi$, for representative values of $\beta$. The crossing of curves corresponding to different values of $\beta$ indicates that under certain conditions, heterogeneity may enable larger quantities of cells or faster dilution rates. This is surprising because larger values of $\beta$ mean that the population of cells concentrate nearer the point of maximum growth rate. The explanation is that in this case larger $\beta$ also leads to higher secretion of toxic byproducts. For example, in Figure \ref{fig:X-D-cho} the red curve has a higher $\beta$ than the black curve, but at $\xi=0.1$ the black curve has higher cell counts. Comparing with Figure \ref{fig:cho-conc}, we see that the red curve at this point also has higher concentrations of the toxic byproducts ammonia and lactate, explaining the reduction in cell growth. This also confirms a prediction already made in the simpler model.

Figure \ref{fig:X-D-cho}(c) plots the steady state values of the cell density for each dilution rate, with unstable steady states shown in dashed lines. As in the simple metabolic model, we find that the system admits more than one steady state for some dilution rates. The dynamical landscape of the system depends on the value of $\beta$. In particular, if $\beta$ is too low the unstable regime disappears, again recapitulating the behavior found in the simple metabolic model.

Since many of the enzymatic flux costs used in these simulations (\emph{cf.} Equation \eqref{eq:flux-costs}) are unknown or poorly annotated, we repeated these simulations after doing random perturbations on all these coefficients of up to 25\% relative to the original values. Figures S3 and S4 in the Appendix show that all of the qualitative features discussed here are preserved under these perturbations.

\section*{Discussion}

In this work we have developed a framework to model cell heterogeneity based on the Maximum Entropy principle (MaxEnt) \cite{jaynes1957information}. In contrast to previous applications of MaxEnt to model metabolic heterogeneity \cite{martino2017statistical}, our work has focused on continuous cell cultures in steady state. As we have shown, in this case it is possible to write down a self-consistent set of equations that determines the distribution of cells in phenotypic space as a function of the dilution rate. The dependence has a non-trivial character due to the nonlinear nature of the chemostat, in some cases including multistable regimes.

We applied this framework in a simple metabolic network first, where all the computations can be carried out exactly. In this toy model we were able to discuss in detail many qualitative features of the model. We found that heterogeneity enables larger populations of cells because of reduced toxicity, that the dynamical landscape includes includes a multistable regime which shrinks or even disappears entirely as the level of heterogeneity increases, and that a byproduct predicted to be zero if heterogeneity is ignored exhibits increasing concentrations when heterogeneity is included in the model.

In this work we have also shown how the Maximum Entropy approach to metabolic modeling can be applied to large metabolic networks, by taking advantage of a recent implementation of the Expectation Propagation algorithm \cite{braunstein2017natcomm}. By exploiting this algorithm we were able to obtain a numerical solution of our model, applied a CHO-K1 genome-scale metabolic network reconstruction. Although this metabolic model is much more complicated than the toy model, it nonetheless recapitulates all the qualitative findings made in the simpler network, but with a richer phenomenology.

In order to obtain a tractable model, we have made some simplifications. We chose to ignore cell-to-cell heterogeneity in mass composition (the parameters $e_i,y_i$ in the notation of Methods). Indeed, although cells can vary widely in their metabolic arrangements, the mass fractions of major constituents such as total protein, lipids, and nucleotides, are approximately invariant for a given cell type. This approximation is not essential for the formal statement of the model, but it greatly simplifies calculations because then the exponent in the Maximum Entropy distribution is a linear function of metabolic fluxes. We have assumed that the availability of nutrients depended only on the total cell concentration. In a more realistic model cells compete for certain metabolites and their concentrations determines the uptake bounds. In the literature models of this type have been studied \cite{mehta2016, swain2017natcomm} and they have an interesting phenomenology of their own. Our approach can be seen as a ``mean-field'' approximation where each cell interacts with the entire population instead of with specific nutrients. We believe that our main conclusions will remain valid if this approximation is relaxed, but further work is needed in this direction. Finally, incorporating the crowding constrain entails replacing reversible reversible reactions by two reactions in the forward and backward direction, almost doubling the total number of reactions. We therefore applied a reduction where reactions found to carry no flux identically for all $\xi$ in the $\beta=\infty$ regime were removed from the model. Obviously this modifies the solution space of the model. On the other hand, an extensive model such as the genome-scale CHO-K1 model \cite{hefzi2016cho} contains many reactions that are inactive under most conditions and therefore including them would only add noise to our analysis. Since FBA has been a successful tool in the detection of relevant metabolic pathways in multiple organisms \cite{shlomi2011ploscb,hefzi2016cho,forster2003gr,Heirendt2017COBRA3,Ibarra2002EColi}, we believe that this reduction improves the quality of our predictions. Indeed, all the byproducts predicted by this model have been observed experimentally to be secreted by mammalian cells \cite{cossio2017ploscb}.

Some general qualitative features are common to both the homogeneous regime ($\beta=\infty$) and the heterogeneous regime (finite $\beta$).
In both cases, multistability is a consequence of negative feedback by toxicity accumulation, although extremeley heterogeneous populations become monostable as shown in Figures \ref{fig:3}, \ref{fig:beta-xi} and \ref{fig:X-D-cho}.
For a given combination of cell-line and feed media, the ratio of cell density to the dilution rate ($\xi = X/D$, inverse of the CSPR \cite{ozturk1996cspr}) determines the steady state of a continuous culture and therefore can be used to connect different modes of continuous cultures (chemostat and perfusion).
This conclusion was obtained in \cite{cossio2017ploscb} in the context of an homogeneous model ($\beta=\infty$).
In the presence of heterogeneity the steady state does not consist of single flux values for every reaction in the network. 
Instead, it must be described by a global probability distribution representing the fraction of cells adopting each metabolic phenotype.
The parameter $\beta$ quantifies the spread of this distribution.
Although following the standard practice of the MaxEnt principle, $\beta$ should be determined to match experimental data on the average growth rate of a population, for the sake of generality, in this work $\beta$ was treated as a free parameter. This is analogous to studying different temperatures in statistical physics.
Therefore the selected values of $\beta$ used in the figures carry no special significance, except that they are representative of the most salient aspects of the general phenomenology of the model.
In the Appendix we show how $\beta$ can be connected to a simple underlying model of the evolutionary dynamics of a cell population.

Interestingly, toxicity also leads to the paradoxical result that more heterogeneous populations can achieve larger sizes.
This happens because at finite $\beta$ less cells are secreting the toxic byproduct at maximal rate.
Thus, a prediction of our model is that inducing heterogoeneity might be beneficial in an industrial setting where cell numbers are limited by the accumulation of toxic byproducts. We also showed that heterogeneity might be responsible for features of the bulk population not derivable from selective pressures on individual cells. The CHO-K1 reconstruction exhibits divergence of formate accumulation, while for the simplified metabolic model it is lactate that accumulates. This difference is not surprising because in the toy model lactate is the only allowed byproduct, while the CHO-K1 model has many possible byproducts. In both cases an homogeneous model would predict zero or constant concentrations.

Compared to population balance models of heterogeneity, our approach has several advantages.
First, by using MaxEnt, assumptions about the detailed mechanisms behind cellular heterogeneity are not required.
As a consequence, the description of heterogeneity is simple and only one additional parameter $\beta$ suffices for this purpose.
This simplicity enables the study of genome scale metabolic networks such as the CHO-K1, which are not accessible to detailed population balance models.

\section*{Acknowledgements}

The authors warmly thank Anna Paola Muntoni for her assistance with the Expectation Propagation code. 





\clearpage

\bibliographystyle{unsrt}
\bibliography{references.bib}

\clearpage

\section*{Supporting Information Legends}
\begin{itemize}
\item{\textbf{S1 Appendix.} Supplementary text}
\item{\textbf{S2 Data.} Reduced CHO-K1 model}
\item{\textbf{S3 Table.} Culture media definition}
\end{itemize}

\end{document}